\documentclass[aps,prl,preprint,superscriptaddress]{revtex4-1}
\usepackage{graphicx,amssymb,amsmath}
\begin{document}


\title{Dynamical Heterogeneity in the Glassy State} 



\author{Apiwat Wisitsorasak}
\affiliation{Center for Theoretical Biological Physics, Rice University, Houston, Texas 77005, USA}
\affiliation{Department of Physics \& Astronomy, Rice University, Houston, Texas 77005, USA}

\author{Peter G. Wolynes}
\email[]{pwolynes@rice.edu}
\affiliation{Center for Theoretical Biological Physics, Rice University, Houston, Texas 77005, USA}
\affiliation{Department of Physics \& Astronomy, Rice University, Houston, Texas 77005, USA}
\affiliation{Department of Chemistry, Rice University, Houston, Texas 77005, USA}


\date{\today}

\begin{abstract}
We compare dynamical heterogeneities in equilibrated supercooled liquids and in the nonequilibrium glassy state within the framework of the random first order transition theory.  Fluctuating mobility generation and transport in the glass are treated by numerically solving stochastic continuum equations for mobility and fictive temperature fields that arise from an extended mode coupling theory containing activated events. Fluctuating spatiotemporal structures in aging and rejuvenating glasses lead to dynamical heterogeneity in glasses with characteristics distinct from those found in the equilibrium supercooled liquid. The non-Gaussian distribution of activation free energies, the stretching exponent $\beta$, and the growth of characteristic lengths are studied along with the four-point correlation function. Asymmetric thermodynamic responses upon heating and cooling are predicted to be the results of the heterogeneity and the out-of-equilibrium behavior of glasses below $T_g$. Numerical results agree with experimental calorimetry. We numerically confirm the prediction of Lubchenko and Wolynes in the glass that the dynamical heterogeneity can lead to noticeably bimodal distributions of local fictive temperatures which explains in a unified way recent experimental observations that have been interpreted as coming from two distinct equilibration mechanisms in glasses.
\end{abstract}

\pacs{}

\maketitle 



%
%

%

\section{Introduction}

The complexity of glass forming systems is most apparent in the unusual nonexponential relaxation  dynamics found both in the supercooled fluid and  in the glassy state. After a supercooled liquid falls out of equilibrium on cooling the resulting glass initially can be thought of as frozen snapshot of the liquid state just before it fell out-of-equilibrium.\cite{wolynes2009spatiotemporal} Thus initially there is no long-range spatial order. Since the glass is out-of-equilibrium, however, it continues to evolve, albeit slowly, as the system ages and new but subtle spatial structure emerges. On experimental timescales ergodicity is broken below the glass transition so time averages no longer equal full-ensemble averages. Furthermore, properties in the glass vary from region to region and the sensitivity of activation barrier heights to local properties leads to very strongly heterogeneous dynamics. In this article we explore the inhomogeneous dynamical structure formed within the glassy state using a framework based on the random first order transition (RFOT) theory.\cite{stoessel:4502,singh1985hard,kirkpatrick1987connections,kirkpatrick1987stable}

The random first order transition theory of glass has already provided a unified quantitative  description of many aspects of supercooled liquids and structural glasses.\cite{lubchenko2004theory} The theory has its origins in bringing together two seemingly distinct mean field theories of glass formation, one being  the dynamical, ideal mode coupling theory,\cite{gotze2009complex} the other being a thermodynamic theory of freezing into aperiodic structures.\cite{kirkpatrick1987connections}  A variety of mean-field like approximations most elegantly formulated via the replica approach deal with the self-generated long-lived randomness in structural glasses.\cite{ParisiZamponi2010MeanField} Within mean field theories there is a special temperature $T_d$, below which an exponentially large number of frozen states emerge.\cite{singh1985hard,kirkpatrick1987stable} It was however easy to see via droplet arguments that beyond the pure mean-field description which gives infinite long lived metastable stats that such many metastable states will be destroyed by locally activated transitions if they are exponentially large in number.\cite{kirkpatrick1987stable,Kirkpatrick1989Scaling} When ergodicity is restored thereby, in finite dimensions, the dynamical transition at $T_d$ is smeared out. Nevertheless at a lower temperature $T_K$ where the configurational entropy of the mean field states would appear to vanish, this argument suggests a true thermodynamic transition could finally appear.\cite{Kauzmann1948EntropyCrisis} Thermally activated motions would cease at such a thermodynamic transition if the entropy crisis occurs rather than being intercepted by some other mechanism. Eastwood and Wolynes have presented an argument that suggests while the droplets themselves could provide such a mechanism of cutting off a transition, on normal time scales, the droplet entropy correction is quantitatively unimportant.\cite{Eastwood2002Droplets}  Droplets provide the key feature of RFOT theory, an activated mechanism of mobility generation. Nevertheless, because of its shared origin with mode coupling theory, RFOT theory allows mobility once generated below $T_d$ to be transported  giving a microscopic basis for the notion of facilitation.\cite{SollichConstrained2003,Garrahan19082003} Facilitation, in its simplest incarnation, can be thought of as defect diffusion, which in low dimensional systems  by itself can lead to nonexponentiality.\cite{Skinner1980Solitons,Skinner1983KineticIsingModel} In the quantitative analysis of non exponential relaxation using the RFOT theory the mobility transport effect was shown to be an essential feature that diminishes the effects of the instantaneous heterogeneity in activation barriers which otherwise  would lead to a broader distribution of relaxation times than observed.\cite{xia2001microscopic}

The entropy driven activated events in RFOT theory allow one to connect a broad range of experimental measurements on dynamics in glasses with their thermodynamics . Examples included the predictions of the fragility index $D$,\cite{xia2000fragilities} the fragility parameter $m$,\cite{stevenson2005thermodynamic} the stretching  exponent in the supercooled liquid $\beta$,\cite{xia2001microscopic} the correlation length\cite{lubchenko2003barrier}  and the yield strength.\cite{Wisitsorasak02102012}  All these predictions require only thermodynamic data as input, while they are all decidedly kinetic observables.

Aside from questions about underlying mechanism even the phenomenological description of dynamics in the glassy state is nontrivial. The nonergodic behavior in the glassy state implies that measured properties not only depend on time but also are dependent in detail on the sample's history of preparation. The need to describe experimental protocols makes even communicating experimental results a challenge let alone predicting results theoretically.\cite{lubchenko2007theory} The fictive temperature concept was introduced as a single additional number globally characterizing the instantaneously non equilibrium state of the glass in the phenomenological description of Narayanaswamy, Moynihan and Tool (NMT). This is not enough. Lubchenko and Wolynes have, however, argued within the RFOT framework that every patch of glass can be described as having a fictive temperature locally.\cite{lubchenko2004theory} In this framework extending the local fictive temperature concept to describe the global level requires dynamically coupling regions together i.e. it requires combining the mobility generation mechanism that acts locally in RFOT with mode coupling that transports mobility. Bhattacharyya, Bagchi, and Wolynes (BBW)\cite{bhattacharyya2005bridging,bhattacharyya2008facilitation,bhattacharyya2007dynamical} have shown to how combine these by including an activated event vertex into the standard MCT vertex. The spatial variations in activated dynamics were not made explicit in their calculations. The theory has been extended to obtain a field theory that takes into account of spatiotemporal structure of dynamics in supercooled liquids and glasses by expanding the mode coupling memory kernels using a Taylor  series in terms of its gradients thus defining a mobility field $\mu$ having explicit space and time dependence.\cite{wolynes2009spatiotemporal} This expansion yields a continuum equation for the mobility field in which mobility can be both generated from activated events or transported from highly mobile regions to less mobile areas through mode coupling facilitation effects. To account for the random nature of mobility generation and transport the  equations for the mobility field also contain generation and transport noises. We have already used these equations to describe the front like transformation of stable glass into supercooled liquids discovered by Ediger \textit{et. al}.\cite{kearns2010observation,sepulveda:204508} The numerical solutions of the continuum equations predict the transformation front speed in quantitative agreement with the experiments.\cite{wisitsorasak2013fluctuating}

While our previous work detailed the spatiotemporal character of the transformation of stable glasses into equilibrium liquids where the mean mobility differs by many orders of magnitude across the sample, even dynamics in the bulk glass alone exhibits heterogeneity in both space and time.\cite{Tracht1998LengthScale,russell2000direct} This heterogeneity is an intrinsic characteristic in  all glassy systems and is evident in non-exponential relaxation.\cite{bohmer1993Nonexponential} The RFOT theory has already explained quantitatively the dynamical heterogeneity in equilibrated suprecooled liquids. Using RFOT theory, Xia and Wolynes argued that due to facilitation the free energy barriers will not be Gaussian distributed, but will have a distribution with significant skewness.\cite{xia2001microscopic} They showed that the non-exponential parameter depends on width of the distribution of the free energy through a static stretching exponent $\beta_0 = 1/\sqrt{1+ (\delta \Delta F/k_B T)^2}$. Here we will show that the continuum equations of a fluctuating mobility field and fluctuating fictive temperature which have been derived  describes more completely the dynamical heterogeneity and directly leads  to a  distribution of the free energy having a long tail on the low energy barrier side as Xia and Wolynes suggested. By following the complete evolution of the mobility field, calculations of the four-point correlation function in the glassy state prove possible.\cite{berthier2007Spontaneous1,berthier2007Spontaneous2}

Dynamical heterogeneity in space and time results in non-trivial out-of-equilibrium signatures in calorimetric experiments which are hysteretic and not symmetric between what happens when glasses are cooled down or are heated up. As explained earlier,\cite{wolynes2009spatiotemporal,lubchenko2004theory} upon cooling, some previously high mobility regions will rapidly equilibrate  to a low fictive temperature near that of the ambient temperature, while other regions, initially more stable, will maintain their higher initial fictive temperature.\cite{wolynes2009spatiotemporal} Since the prematurely equilibrated regions now have a low mobility one should find in the partially aged glass a bimodal rate distribution. Mobility transport to the low fictive temperature regions also slows approach to equilibrium in their neighborhoods. In contrast, upon heating, the glasses after the first reconfiguration events, regions near to the first reconfigurating regions equilibrate faster than they would have in the original equilibrated sample. The initially reconfigured regions of high mobility now catalyze the rearrangement of their low-mobility neighbors leading to a radially propagating front of mobility. The numerical solution of the continuum equations of mobility field and the fictive temperature that we are going to present in the next section  quantitatively captures these predicted phenomena. 


The organization of this article is as follows. In section II we provide a more explicit derivation of the continuum equations for the mobility field driven by activated events with mobility fluctuations than was given before. We then explore the inhomogeneous dynamical structure of the glass in section III where we analyze the free-energy distribution, the four-point correlation function, and the stretching exponent for $\alpha$-relaxation. In section IV, we use the equations to model calorimetric experiments on several glasses and compare predictions with experimental data. In section V, we investigate the kinetics of long-aged glasses and show that the bimodal dynamic heterogeneity which emerges will appear in the laboratory as ``ultraslow'' relaxations like those experimentally observed\cite{Miller1997Ultraslow} and recently highlighted.\cite{Cangialosi2013TwoEquilibration} Section IV summarizes our study and discusses possible extensions of this work. 

\section{The continuum equations of mobility and fictive temperature} \label{sec:continuumequations}
In the following we describe in some detail the microscopic basis of the continuum equations of mobility field with fluctuating generation and transport described earlier by one of us.\cite{wolynes2009spatiotemporal} These equations were previously studied numerically by us\cite{wisitsorasak2013fluctuating} in the context of mobility front propagation in which an ultrastable glass prepared by vapor deposition ``melts'' into an equilibrated liquid from the free surfaces which are predicted by RFOT theory to have higher mobility than the bulk. RFOT theory near an interface predicts $\tau_{surface} = \sqrt{\tau_0\tau_{bulk}}$ where $\tau_0$ is the microscopic time, $\tau_{bulk}$ is the relaxation in the three dimensional bulk and $\tau_{surface}$ is the relaxation time of the surface layer.\cite{stevenson2008surface}

In the equilibrium supercooled liquid, particles spend relatively brief times near each other, collide and then move freely and go on further to collide with others in a weakly correlated way. As the temperature decreases or the density increases further, groups of correlated collisions between particles in the same spatial region occur more and more frequently. Persistent structures therefore emerge. Individual particles continue to move and vibrate, but now reside within a nearly fixed cage. Once in awhile, particles nevertheless manage to move from one cage to another through activated processes which are actually many particle events. Thus confined motion followed by hops becomes the dominant dynamical feature at temperatures below $T_d$. Extending mode coupling theory through more elaborate perturbative expansions that only keep track of a finite number of correlated collisions will miss the activated process which is an essential singularity in the temperature (and therefore fluctuation strength) expansion. This is why any simple perturbation extension at finite order in MCT or in static replica methods is problematic, at best. This formal difficulty exists also in the theory of ordinary first order transitions\cite{Binder1987FirstOrder} and in the quantum chromodynamics of the nucleon.\cite{Schafer1998QCD} 

The standard mode-coupling theory gives an equation of motion for the system average correlation function that decays from a constant value to zero in the long time limit above the dynamical transition temperature. The decay of the correlations depends on the history of the system in the past. The nature of the decay is mathematically encoded in the memory kernel.\cite{zwanzig1965time} This kernel can be expanded perturbatively in terms of higher order correlation functions, but can be approximated self-consistently as products of two-point correlation functions.\cite{gotze2009complex} The resummed perturbative mode coupling theory works perfectly well at temperatures, above $T_d$,\cite{Reichman2005MCT} but the theory breaks down at the temperature below $T_d$ as it shows an infinite plateau in the two-point correlation function,\cite{Mezei1999Intermediate} which is not allowed in finite dimension above $T_K$, i.e. the cages are permanent in mode coupling theory.  Instead of complete freezing the system actually goes from collision dominated transport to transport driven by activated processes as it crosses the dynamical temperature $T_d$.\cite{lubchenko2007theory} 

Being essential singularities the activated dynamics must be incorporated  into the MCT directly. As shown by BBW,\cite{bhattacharyya2005bridging} activated processes give an extra decay channel for the correlation function. When the activated events are included, the memory kernel consists of two distinct contributions one coming from idealized MCT kernel $M_{mct}(k;t,t')$ and one described by an additional hopping kernel $M_{hop}(k;t,t')$
\begin{eqnarray}
M_{total}(k;t,t') = M_{mct}(k;t,t') + M_{hop}(k;t,t').
\end{eqnarray}
The mode coupling memory kernel has a complicated structure involving coupling to density fluctuation modes with other wave vectors, but it  is dominated by density modes with wave vector $k$ near the value corresponding to the first peak of the structure factor, since Fourier density modes of the wavelength decay the slowest due to de Gennes narrowing.\cite{DeGennes1959Narrowing} This simplification leads to the so-called schematic mode coupling theory. 

A mobility field can be defined as the longtime rate at which particles reconfigure through activated dynamics. This field can thus be formulated in real space as a Fourier transform in space of that memory kernel and is an integral of that kernel over elapsed time $t'$.
\begin{eqnarray}
\mu(r,t) = \int_{-\infty}^tdt'  \int_{-\infty}^{\infty} dk e^{-i k \cdot r} M_{total}(k;t,t'). \nonumber
\end{eqnarray}
The total mobility field will contain two components one from the activated dynamics and one from the idealized MCT part
\begin{eqnarray}
\mu(r,t) = \mu_{hop}(r,t) + \mu_{mct}(r,t),
\end{eqnarray}
where
\begin{eqnarray}
\mu_{hop}(r,t) = \int^t dt' \int_{-\infty}^{\infty} dk e^{-i k \cdot r} M_{hop}(k;t,t') \nonumber
\end{eqnarray}
and
\begin{eqnarray}
\mu_{mct}(r,t) & = & \int^t dt' \int_{-\infty}^{\infty} dk e^{-i k \cdot r} M_{total}(k;t,t') \nonumber \\
\ & = & \int^t dt' M_{mct} \left( x+ \frac{\delta R}{2},x - \frac{\delta R}{2};t,t' \right). \nonumber
\end{eqnarray}
We assume that the part of the mobility coming from hopping dynamics $\mu_{hop}$ instantaneously re-adjusts to depend on the local fictive temperature $T_{f}$ and the ambient temperature $T$. The MCT memory kernel is determined by the full dynamic density correlation function $\phi(k,t) \approx \phi_{MCT}(k,t) \phi_{hop}(k,t)$,
\begin{eqnarray}
M_{mct}(r,r';t,t')  \sim \int dk \int dk_1 e^{-i k \cdot (r-r')} V(k,k_1;t,t') \cdot \phi(k;t') \phi( \vert k - k_1 \vert;t'),
\end{eqnarray}
where $V(k;t,t')$ is the usual vertex function in the standard MCT. 

Ultimately the mobility field differs from one region to another because the local mobility has inherited local fluctuating structure from the liquid state. Particles residing in the cage also have different activated relaxation times depending on the local fictive temperature which itself varies in space and time. These effects both lead to the mobility being inhomogeneous in space and time.  In the approximation that these quantities do not change too rapidly from one place to neighboring regions, we can expand the mobility field in a Taylor's series around $x$ and $t$. Doing this leads to an equation of motion for $\mu_{mct}$. 
To see this, note the correlation function $\phi(k,t)$ itself depends on the memory kernel which again varies in space and time because mobility does. Now if we expand $\phi \left(k,\mu(x',t') \right)$ about $\mu(x,t)$ we find that
\begin{widetext}
\begin{eqnarray*}
\phi(k,\mu(x',t')) &\approx& \phi \left(k,\mu(x,t) + (x'-x) \nabla \mu + \frac{(x'-x)^2}{2} \nabla^2 \mu(x,t) + (t'-t) \frac{\partial \mu}{\partial t} + \ldots \right)  \\
 & \approx & \phi(k,\mu(x,t)) + \frac{\partial \phi}{\partial \mu} \left((x'-x) \frac{\partial \mu}{\partial x} + (t'-t) \frac{\partial \mu}{\partial t} + \ldots \right) + \ldots \\
 &\approx& \phi(k,\mu(x,t)) + \frac{\partial \phi(k,\mu(x,t)}{\partial \mu} \Delta \mu + \frac{1}{2} \frac{\partial^2 \phi}{ \partial \mu^2} \Delta \mu^2 + \ldots 
\end{eqnarray*}
where $\displaystyle \Delta \mu \equiv \mu(x',t') - \mu(x,t)  = (x'-x) \nabla \mu + (t'-t) \frac{\partial \mu}{\partial t} + \mathcal{O}\left(\nabla^2 \mu, \partial^2 \mu/ \partial t^2 \right)$.
\end{widetext}

Now we can expand $\phi(x,\mu(t))$ and keep the lowest terms in the gradients. The gradient expansion of $\mu_{mct}$ thus has the form:
\begin{eqnarray}
\mu_{mct}(x,t) & = & \bar \mu_{mct}(x,t) + w_0 \frac{\partial \mu_{mct}}{\partial t} \nonumber \\  
& \ & + w_1(\nabla^2 \mu_{mct}) + w_2 (\nabla \mu_{mct})^2 
\end{eqnarray}
where $\displaystyle \bar \mu_{mct} = \bar \mu - \mu_{hop} = (1-\lambda) \bar \mu$. $\bar \mu$ is a locally equilibrated mobility  field which we will discuss later. The parameter $\lambda$ is the ratio of the part of the mobility field  coming from the activated events alone to the total mobility. This quantity depends on the local structure factors and has been computed for \textit{Salol} by BBW.\cite{bhattacharyya2008facilitation} For salol they find $\mu_{hop} = \lambda \bar \mu \approx 0.25 \bar \mu$. For convenience we will assume in our numerics that this factor $\lambda=0.25$ is the same for all glasses. The coefficients $w_0,w_1$ and $w_2$ are  also given by integrals over the MCT kernel. For example the coefficient $w_0$ can be written explicitly as
$$
w_0 = \int\limits_{-\infty}^0  V(\delta R=0,t,t') \cdot \left[ \phi_k \frac{\partial \phi_p}{\partial \mu} + \phi_p \frac{\partial \phi_k}{\partial \mu} \right] (t-t') d(t-t'),
$$
where $V(\delta R = 0,t;t')$ is the vertex function of standard mode-coupling theory.\cite{gotze2009complex}

Notice if we approximate each term of $\phi$ by its long time behavior we can write $\displaystyle \phi \approx \phi_0 e^{- \int^{t-t'}_{\infty} \mu d\tau} \approx \phi_0 e^{- \mu (t-t')}$, so then we obtain
\begin{equation}
\frac{\partial \phi}{\partial \mu} = -(t-t') \phi(t - t') \nonumber
\end{equation}
Carrying out the integral, we then get a simple form for the coefficient $w_0 = -\frac{1}{2 \mu^2} \cdot \bar \mu_{mct}$.

In the absence of spatial inhomogeneity, the equation of motion for $\mu_{mct}$ is
\begin{equation}
\mu_{mct} - \bar \mu_{mct} = -\frac{\bar \mu_{mct}}{2 \mu^2}\cdot\frac{\partial \mu_{mct}}{\partial t}
\end{equation}

Now we examine the other gradient terms, within each term of the Taylor series, a term of the type $ \partial \phi /\partial \mu$ or $\left(\partial \phi / \partial \mu \right)^2$ is generated. In each term again using the exponential approximation for the local decay gives for the derivative  $(t-t') \phi(t-t')$. Thus within this approximation one generates terms of the same form as in the MCT kernel but now with extra powers of $(t-t')$. Upon integrating then one finds extra factors $\frac{1}{\mu^K} \cdot K!$ for a term containing $(t-t')^K$. Except for powers of $\delta R^N$ one again gets the mode coupling kernel. Thus a typical term will be $\frac{K!}{\mu^K} \int d\delta R \int\delta R^N M \left(x-\frac{\delta R}{2}, x+\frac{\delta R}{2}, t-t' \right)d(t-t')$. With the spatial variation, the equation of motion for $\mu_{mct}$ also has spatial gradient terms:
\begin{widetext}
\begin{eqnarray*}
\mu_{mct} - \bar \mu_{mct} &=& -\frac{\bar \mu_{mct}}{\mu_{mct}^2} \cdot \frac{\partial \mu_{mct}}{\partial t} \\
& \ & + \kappa(\nabla \nabla \mu_{mct})\cdot \frac{1}{\mu_{mct}} \int d \delta R\ dt \delta R^2 M\left(x+\frac{\delta R}{2},x - \frac{\delta R}{2} \right) \\
& \ & + (\nabla \mu_{mct})^2 \cdot \frac{1}{\mu_{mct}^2} (\mbox{similar integral})
\end{eqnarray*}
\end{widetext}

Notice that a characteristic length $\xi$ naturally emerges from the ratio of the two integrations in the gradient term versus uniform MCT:
\begin{equation}
\frac{\int \delta R^2\ M dt\ d\delta R}{\int M dt \ d\delta R} \sim \xi ^2
\end{equation}
So the coefficient of the $\nabla \nabla \mu_{mct}$ can be written as $\xi^2 \mu_{mct}$. Thus we see that $\xi^2\frac{\mu_{mct}^2}{\bar \mu_{mct}}$ can be thought of as the mobility diffusion coefficient. Since activated events are included in the kernels the $\xi$ obtained in this way will not precisely be equal to the correlation length of the pure mode coupling theory $ \Xi_{MCT}$ which was determined by Reichman \textit{et. al.},\cite{Reichman2005MCT} but it will be related to it sufficiently far from $T_d$ itself.

By taking the static limit we can see that the coefficient of the linear Laplacian term contains a length $\xi$ which would be the correlation length of the 4-point correlation function consistent with an analysis by Biroli \textit{et. al.}\cite{biroli2006inhomogeneous} The gradient squared term arises because of the nonlinear relation of the MCT closure between the memory kernel and the correlation functions. The coefficients $\xi^2$ and $w_2$ depend on the details of the microscopic mode coupling closures employed. For simplicity we will choose the value of $w_2$ so that the locally linearized equation can be written as a mobility flow equation with a source term
\begin{eqnarray} \label{eq:MuMCT}
\frac{\partial \mu_{mct}}{\partial t} & = &\frac{\partial }{\partial x} \left(\frac{2 \mu^2 \xi^2}{\bar \mu_{mct}} \frac{\partial \mu_{mct}}{\partial x} \right) \nonumber \\
& \ & -\frac{2 \mu^2}{\bar \mu_{mct}} \left( \mu_{mct} - \bar \mu_{mct}\right)
\end{eqnarray}
where $\mu_{mct} = \bar \mu - \mu_{hop}$. We believe that apart from this bulk source term any violation of the  local ``conservation'' laws for mobility will be small and in any case one can see they would only modify the results in a modest quantitative fashion.

The uniform solution of the extended mode coupling theory with activated events $\bar \mu_{mct}$ which is used for input can be approximated by using the ultralocal theoretical analysis of aging glasses by Lubchenko and Wolynes.\cite{lubchenko2004theory} They suggest activation will still be the dominant effect below $T_d$ so that the rates depend on the local fictive temperature $T_f$ and ambient temperature $T$
\begin{eqnarray}
\bar \mu = \mu_0 \exp \left\{ - \frac{\gamma^2}{4 k_B T \Delta c_p Tg \left( \frac{T-T_K}{T_K} - \ln{\frac{T}{T_f}} \right) } \right\}
\end{eqnarray}
where $\displaystyle \gamma = \frac{3 \sqrt{3 \pi}}{2} k_B T_K \ln \left[ \frac{(a/d_L)^2}{\pi e} \right]$, $d_L$ is the Lindemann length and $a$ is the interparticle spacing. The details of their derivation can be found in Appendix \ref{app:appendix1}.

The fictive temperature approximately obeys an ultralocal relaxation law with the local decay rate $\mu(r,t)$.
\begin{eqnarray}\label{eq:Tf}
\frac{\partial T_f}{\partial t} = - \mu \left(T_f - T \right).
\end{eqnarray}

In the bulk itself there are dynamic inhomogeneities from fluctuations in the local fictive temperature and from the happenstance nature of the  activated events which generate and transport the mobility field. We treat the corresponding random force terms in the equation as coarse-grained white noises with strengths and correlation lengths that reflect the length scale of the activated events. The intensities of the noises may be found by requiring the linearized equations to satisfy locally fluctuation-dissipation relations. The local fictive temperature fluctuation $\delta T_f$ is taken to be $\displaystyle \langle \delta T_f (x,t) \delta T_f(x',t') \rangle = 2 \mu T^2 \frac{k_B}{\Delta c_p N^\ddag} \delta (x-x') \delta (t-t')$, where $N^\ddag$ is the number of molecular units in a cooperatively rearranging region. The fluctuations in mobility generation $\delta g$ and transport $\delta j$ arise due to fluctuations in free energy barrier heights which are the primary cause of the stretched exponential relaxation with a bare stretching parameter $\beta_0 = 1/\sqrt{1+ (\delta F^\ddag/k_B T)^2}$. Linearizing the mobility equation and treating $\bar \mu$ as constant, the local fluctuation-dissipation relation yields a mobility generating noise with correlations  $\langle \delta g(x,t) \delta g(x',t') \rangle = 2 \bar \mu \mu^2 \left(1/\beta_0^2 -1 \right) \left( T_g/T\right)^2 \delta (x-x') \delta (t-t')$ and a mobility random flux with correlations $\langle \delta j(x,t) \delta j(x',t') \rangle = 2 \bar \mu  \xi^2 \mu^2 \left(1/\beta_0^2 -1 \right) \left( T_g/T\right)^2 \delta (x-x') \delta (t-t')$.\cite{wisitsorasak2013fluctuating}

\section{Inhomogeneous dynamical susceptibilities}

Both the equilibrated liquid and the nonequilibrium glass have intrinsically disordered structures. Even if all particles are identical, in each system, individual particles experience different local environments at any given time. In the high temperature liquid, the disorder may be averaged over and the system described via uniform equilibrium. But, as the glass transition is approached, the structural relaxation time increases dramatically and the system takes a long time to relax and may find itself out-of-equilibrium.\cite{Cugliandolo1994OutOfEquilibrium,lubchenko2004theory} Disorder in the glass seems to be static yet molecules are in motion and change their locations through activated dynamics.\cite{Gruebele2010DirectVisualization} Some particles may move large distances, while others remain localized near their original position. We may refer these behaviors as particles being ``mobile'' or ``immobile'', and any given sites may be characterized by quantity describing themselves the so-called ``mobility.''\cite{berthier2011overview} Experiments show that the mobility of particles in adjacent regions can vary by several orders of magnitude.\cite{russell2000direct}

\begin{figure}
\includegraphics[width=19pc,angle=0]{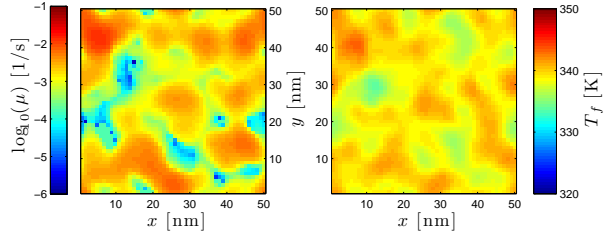} 
\caption{ (Color online) Snapshots of the mobility field in \textit{log} scale (left) and the fictive temperature field (right). \label{fig:mobilityfield1}}
\end{figure}

Over the last decade, direct molecular dynamics simulations of supercooled liquids have given some insight into the nature of dynamic heterogeneity \cite{berthier2011dynamical} but it has been a challenge to go further into deeply low temperature region characteristic of laboratory experiments because of diverging relaxation time. Here we can study realistic levels of supercooling by using the methods described in the previous section to simulate the heterogeneity of the glass  starting from the mesoscopic equations for the mobility field. A snapshot of the mobility and the fictive temperature field of 1,3,5-Tris(naphthyl)benzene (TNB) glass which has been equilibrated at $T=T_g=347$ K by solving the fluctuating field equations in two dimensions is shown in figure \ref{fig:mobilityfield1}. The highly mobile regions of particles are colored in red and the low mobility regions are colored in blue. As seen in the figure, the mobility field varies throughout the space in a way that correlates with the local values of the fictive temperature. The high mobility regions (low relaxation time) are generally less stable than the regions with low mobility and are ready to reconfigure to lower energy state. The fictive temperature and the mobility are coupled in a non-linear fashion; however, their values are consistent with each other in general. The low mobility regions generally map on to the low fictive temperature and vice versa. In equilibrium both fields fluctuate around their equilibrium values. 

\subsection{Activation Free Energy Barrier Distribution}
\begin{figure}
\includegraphics[width=19pc,angle=0]{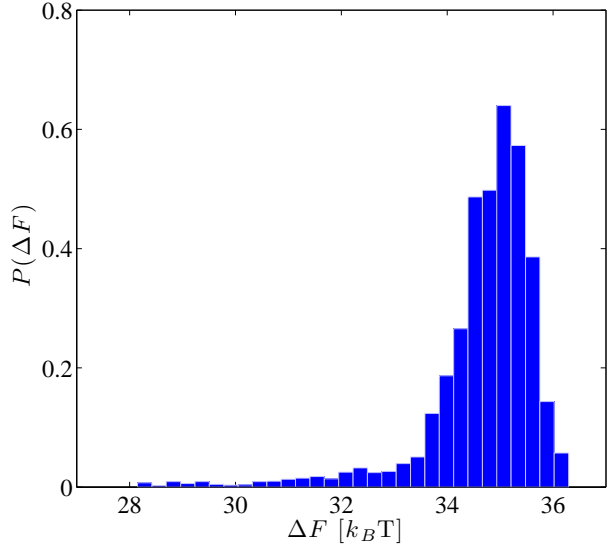} 
\caption{ (Color online) Probability distribution of activation free energy from the numerical simulation of the mobility field and the fictive temperature. Notice that the distribution has a long tail on the low energy barrier side and its shape is not Gaussian distribution. \label{fig:energy distribution}}
\end{figure}

It is an important point to mention that the domains surrounding any region may reconfigure before that central region can move. When this happens local constraints on the slow region will be removed. This change of environment effect can be called ``facilitation.'' As Xia and Wolynes argued this effect means the static barrier height distribution coming from fictive temperature fluctuating will be cut off on the high barrier side. A simple cutoff distribution follows from  the idea that it is primarily those domains slower than the most probable rate that would actually reconfigure only when their environmental neighbors have changed; thus they actually will reconfigure at something close to the most probable rate, which has already been predicted by the RFOT theory. The resulting cutoff distribution for activation barrier, as discussed by Xia and Wolynes,\cite{xia2001microscopic} has a form
\begin{eqnarray}
P(\Delta F^\ddag) = \left\{ \begin{array}{l l} 
P_f(\Delta F^\ddag), & \text{for} \Delta F^\ddag < \Delta F^\ddag_0 \\
C \delta(\Delta F^\ddag - \Delta F^\ddag_0).  & \
\end{array} \right. \label{eq:XWenergycutoff}
\end{eqnarray}
The delta function is an exaggeration of the effect as pointed out by Lubchenko who has provide a more elegant distribution.\cite{Lubchenko2007ChargeMomentum} The free energy barrier found from the simulation does not contain a sharp delta function in Eq.~(\ref{eq:XWenergycutoff}). Figure \ref{fig:energy distribution} illustrates the free energy distribution from the simulation. Like the Xia-Wolynes distribution\cite{xia2001microscopic} and Lubchenko's,\cite{Lubchenko2007ChargeMomentum} the shape of the distribution is not symmetric and has a long tail to the low energy barrier side. This result reflects that the mobility can be transported from higher region to lower region through the facilitation effect.

It should be noted that because of the Arrhenius law the free energy barriers being even modestly distributed in space leads to the wide variation of the relaxation times. Each domain relaxes independently with something close to a  simple exponential decay function. By virtue of of the ensemble average done in most bulk experiments, however, the resulting bulk relaxation exhibits a highly non-exponential distribution of times for the whole sample $\phi (t) \sim e^{-\left( t / \tau \right)^{\beta}}$. That this is the primary origin of dynamical heterogeneity in glasses and other disordered material has been emphasized by several studies.\cite{Richert1993OriginDispersion,richert1994disorder,Ediger2000Heterogeneous} Defect diffusion in three dimensional systems is a secondary effect.

\subsection{Nonexponentiality of relaxation below $T_g$}

As the supercooled liquid falls out of equilibrium, particles can be considered to form groups of cooperatively rearranging regions. This mosaic structure of the liquid gives rise to dynamical  heterogeneity and the overall non exponential relaxation. The driving force for reequilibration varies from mosaic cell to cell. This leads to a wide range of activation barriers $\Delta F^\ddag$.  Computing these fluctuations allowed Xia and Wolynes to predict $\beta$ for a range of substances in the equilibrated liquid.\cite{xia2001microscopic} Their approach was extended to the aging regime by Lubchenko and Wolynes.\cite{lubchenko2004theory}

\begin{figure}
\includegraphics[width=19pc,angle=0]{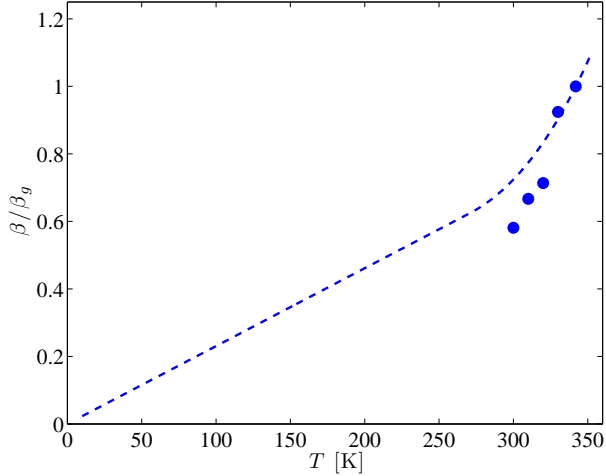}%
\caption{ (Color online) The variation with temperature of the stretch exponent $\beta$ for TNB glass in comparison with $\beta (T_g)$. The predicted value from the static RFOT is shown as a dashed line. The numerical result (solid circle) are in agreement with the RFOT prediction (dashed line). \label{fig:betaratio}}
\end{figure}

Below the glass transition temperature $T_g$, the local structure of the liquid is initially frozen at $T_g$. The ratio of the size of the $\Delta F^\ddag$ fluctuations in the frozen, cooled state, $\delta \Delta F^\ddag_{n.e.}$ to that found at $T_g$, $\delta \Delta F_g^\ddag$ is predicted by Lubchenko and Wolynes to be: 
\begin{eqnarray}
\frac{\delta \Delta F^\ddag_{n.e.}}{\delta \Delta F^\ddag_g} = \frac{\Delta F_{n.e.}^\ddag}{\Delta F_g^\ddag} \frac{T_g s_c(T_g)}{\phi_{in} - \phi_{eq} + T s_c(T)}.
\end{eqnarray}
For a strong liquid, $\Delta c_p$ is small so that there  are small energy fluctuations leading to small $\delta \Delta F^\ddag$. Thus $\beta$ remains near to one until very low temperature for strong liquids. For the fragile liquid, however the fluctuations are very large,  so we would find
\begin{eqnarray}
\beta \approx \frac{k_B T}{\delta \Delta F^\ddag}.
\end{eqnarray}
Lubchenko and Wolynes thus predicted the ratio of the stretching exponent in the initially formed non-equilibrium regime to the stretching exponent at $T_g$ to be: 
\begin{eqnarray}
\frac{\beta_{n.e.}(T)}{\beta(T_g)} = \frac{T}{T_g} \left[\frac{\Delta F^\ddag_{n.e.}(T)}{\Delta F^\ddag_g} \right]^{-2} \left[\frac{\gamma(T)}{\gamma(T_g)} \right]^{-2}.
\end{eqnarray}
This ratio predicted by Lubchenko and Wolynes is plotted as a dashed line in figure~\ref{fig:betaratio} along with the ratio obtained from the present numerical simulation of the TNB glass. As we can see, both simulation and the earlier RFOT estimation for the ratio agree quite well. Unfortunately carrying out simulations further below $T_g$ where a change in slope of the $\beta$ versus $T$ relation is predicted takes a very long time, so we do not have a comparison for the slope change at the present.

\subsection{Length Scales of Dynamical Heterogeneity from Four-Point Correlation Functions}

In this section, using the simulation results developed in section~\ref{sec:continuumequations}, we can study the evolution of the spatial character of the fluctuations below $T_g$. We focus, in particular, on the four-point correlation function used to probe dynamical heterogeneity and show that the four-point susceptibility $\chi_r(t)$ increases as the temperature decreased.

The four-point correlation function was first studied in molecular dynamic simulations of a Lennard-Jones mixture by Dasgupta \textit{et al.} \cite{Dasgupta1991GrowingCorr} The motivation of their work was to demonstrate the presence of a growing length scale associated with fluctuations of the Edwards-Anderson order parameter. We show in this section that the growth of the four-point susceptibility follows naturally from mobility fluctuations with kinetic inputs from material specific thermodynamic properties.

The mobility $\mu$ indicates how long a particle $i$ takes to reconfigure since its last reconfiguration event. Thus a local two-point correlation function can be defined as
\begin{eqnarray}
C(r;t_0,t) \equiv \exp \left( \int_{t_0}^t dt' \mu(r,t') \right).
\end{eqnarray}

The spatial fluctuation of this local two-point correlation function then naturally leads to the four-point correlation function
\begin{eqnarray}
G_4(r;t) & \equiv & \left\langle \int dr' C(r'; t_0, t) C(r' + r; t_0, t) \right\rangle \nonumber \\
& \ &  - \left\langle \int dr' C(r';t_0,t) \right\rangle^2.
\end{eqnarray}

We can define a susceptibility associated with this correlation function
\begin{eqnarray}
\chi_4(t) \equiv \int dr G_4(r;t).
\end{eqnarray}

\begin{figure}
\includegraphics[width=19pc,angle=0]{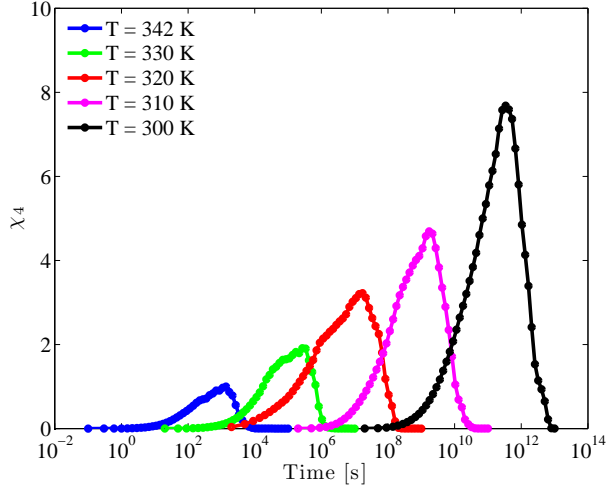}%
\caption{ (Color online) The time and temperature dependence of $\chi_4(t)$ for the TNB glass from the simulation. For each temperature, $\chi_4(t)$ has a maximum, which shifts to larger times and has a larger value when $T$ is decreased, revealing the increasing length scale of dynamic heterogeneity.  \label{fig:chi4}}
\end{figure}

The function $\chi_4(t)$ has been computed from our numerical field theoretic simulations of TNB glasses and is shown in figure~\ref{fig:chi4}. The qualitative course of the temporal behavior is the same at  all the different temperatures: as a function of time $\chi_4(t)$ first increases, then has a peak on a time scale that tracks the structural relaxation time scale and then finally decreases.

\begin{figure}
\includegraphics[width=19pc,angle=0]{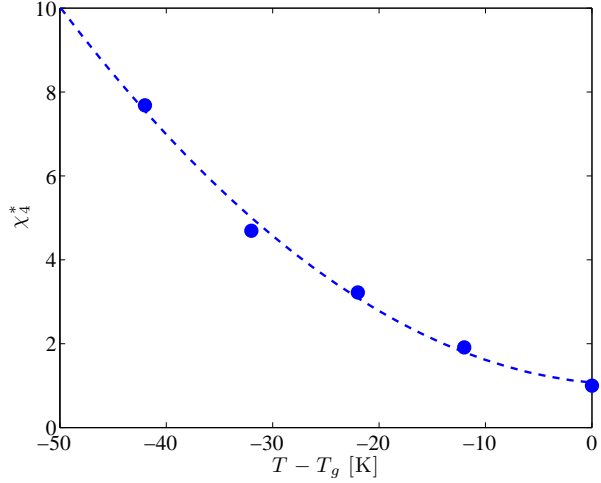}%
\caption{ (Color online) The maximum value of $\chi_4(t)$ of TNB glass as a function of temperature. \label{fig:maxchi4}}
\end{figure}

The peak value of $\chi_4(t)$ measures the volume over which the structural relaxation processes are correlated. Figure~\ref{fig:maxchi4} presents the temperature evolution of the peak height, which is found to increase when the temperature decreases and the dynamics slows down. From such data, one sees directly that increasingly long-ranged spatial correlations emerge as the temperature is lowered. This conclusion supports the idea that at low temperatures the growing length scale calculated from RFOT theory is what should actually appear in the dynamical four point function. The detailed calculation of the length scale at temperature below $T_g$ is shown in appendix~\ref{app:corrlength}. 

\begin{figure}
\includegraphics[width=19pc,angle=0]{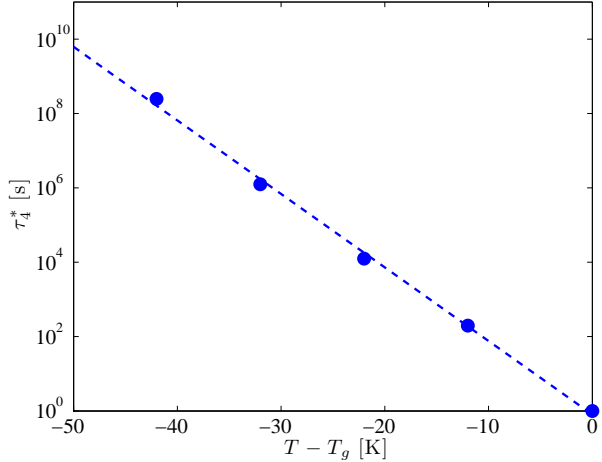}%
\caption{ (Color online) The corresponding time at the maximum value of $\chi_4(t)$ of TNB glass as a function of temperature.  \label{fig:timemaxchi4}}
\end{figure}

Figure~\ref{fig:timemaxchi4} shows the time corresponding to the maximum value of $\chi_4(t)$. The behavior of $\tau^*$ is similar to the structural relaxation time which increases as the temperature is decreased. 

\section{Calorimetry}
\subsection{Introduction}

As the glass transition is approached by cooling, the relaxation time of the supercooled liquid increases significantly. At the glass transition temperature itself, the inverse of the normalized cooling rate $d \log T/ d t$ is approximately equal to the relaxation time. As the cooling process continues, the structural  relaxation of the liquid cannot catch up with the changing ambient temperature so the system remains out-of-equilibrium and becomes a glass. Just below the glass transition temperature, the system freezes into the amorphous state with the structure of supercooled liquid at $T_g$. But small changes continue to occur as the system continues to cool and these changes can be monitored by calorimetry. These changes can then be undone by heating the sample. We now study with our numerical solution an idealized calorimetric experiment based on simple cooling and heating protocols. Usually calorimetry experiments measure the (vibrational) temperature $T$ as a function of time. The temperature is lowered with the cooling rate $\nu_c \equiv \Delta T/ \Delta t <0$. Then the temperature is raised with the heating rate $\nu_h \equiv \Delta T/\Delta t >0$. Conventionally the glass transition temperature is the mid-point temperature on the cooling scan where the heat capacity changes from that of the liquid state to that characteristic of the amorphous state. This kind of experiment is often called differential scanning calorimetry (DSC). DSC is a standard technique for determining the glass transition temperature $T_g$ of noncrystalline materials such as supercooled organic liquids, metallic glasses.\cite{richert2011HeatCapacity} This technique measures  the difference in the amount of heat required to increase or decrease the apparent temperature of a sample as a function of temperature. In an equilibrium liquid or solid, both the sample and the reference thermometer are maintained at nearly the same vibrational temperature.

The DSC experiment thus determines a nonequilibrium heat capacity, or more precisely said, the rate of change of enthalpy as a function of time. To a good approximation, the equilibrium density of configurational enthalpy is a linear function of temperature over the relevant range of temperatures. With this approximation, the non equilibrium configurational heat capacity is $C_p = C_p^{glass} + \Delta C_p \nu^{-1} dT_f/dt$, that is, 
\begin{eqnarray}
\tilde C_p(T) \equiv [C_p(T) - C_p^{glass}]/\Delta C_p = dT_f / dT,
\end{eqnarray}
where $\Delta C_p$ is the difference between liquid and glass heat capacities outside the glass-transition region and $T$ refers to the ambient temperature at the time when the non equilibrium heat capacity is measured. The (global) fictive temperature, has been defined as the temperature at which the properties of a vitreous system in a given state are equal to those of the equilibrium liquid. Locally fictive temperature is not uniform but varies following the ultra local relaxation law with the rate depending on the local mobility field $\mu(x,t)$,
\begin{eqnarray} \label{eq:Tfeqn}
\frac{dT_f}{dt} = - \mu \left( T_f - T \right) + \delta \eta.
\end{eqnarray}
The local mobility field is explicitly determined by the continuum equation that we derived earlier and supplemented by fluctuating mobility generation and transport
\begin{eqnarray} \label{eq:MuMCTFluc}
\mu & = &  \mu_{hop} + \mu_{mct} \nonumber \\
\frac{\partial \mu_{mct}}{\partial t} & = &\frac{\partial }{\partial x} \left(\frac{2 \mu^2 \xi^2}{\bar \mu_{mct}} \frac{\partial \mu_{mct}}{\partial x} \right) \nonumber \\
& \ & -\frac{2 \mu^2}{\bar \mu_{mct}} \left( \mu_{mct} - \bar \mu_{mct}\right) + \delta g + \frac{\partial \delta j}{\partial x}
\end{eqnarray}
where $\bar \mu_{mct} = \bar \mu - \mu_{hop}$.

Below the glass transition temperature $T_g$ the structural relaxation time $1/\mu$ is longer than time required to lower the ambient temperature so that the fictive temperature $T_f$ is approximately fixed at $T_g$. In contrast, above $T_g$, $T_f$ is close to the actual ambient temperature. 

\subsection{Results: Calorimetry at Constant Heating and Cooling Rate}

\begin{figure}
\includegraphics[width=19pc,angle=0]{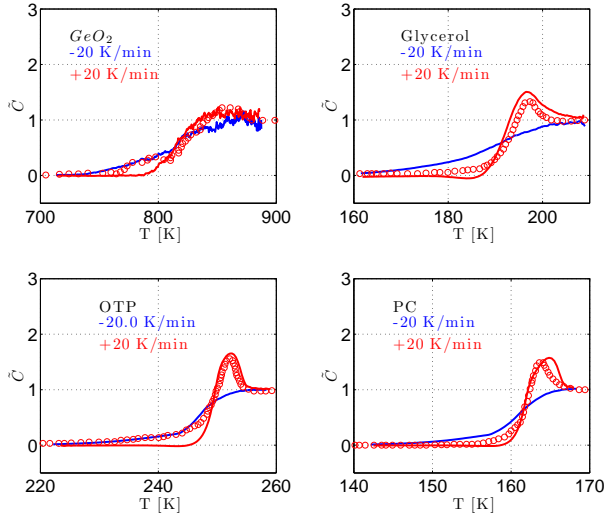}%
\caption{ (Color online) Experimental data are depicted by red circles for heating scan. Red and blue lines represent heating and cooling curves obtained from our calculation. \label{fig:HeatCapacityFourSubstances}}
\end{figure}

Figure~\ref{fig:HeatCapacityFourSubstances} presents the both heating and cooling $\tilde C_p$ curves for $\mathrm{GeO_2}$, Glycerol, \textit{O}-terphenyl (OTP), and propylene carbonate (PC) and those found by solving the numerical equation of MCT/RFOT theory. These substances were chosen because they cover a wide range of fragility, ranging from strong to fragile glasses. The standard scans are considered with the cooling and heating rates equal to 20 K/min as in experiment. The graphs include both the experimental data\cite{Velikov2001CalorimetricOTP,wang2002ScanningCalorimetry,Hu2008SecondRelaxation,Richert2011HeatCapacityPC} (circles) and the simulation results (solid lines). The corresponding cooling  and heating curves obtained from our model are depicted by blue and red lines, respectively. The solid lines in these plots are the results of numerically solving the coupled equations of the mobility field and the fictive temperature following Eq.~(\ref{eq:MuMCTFluc}) and Eq.~(\ref{eq:Tfeqn}). Overall our simulation results show excellent quantitative agreement with the experiments; however, some small deviations between our simulations and the experiments might be seen. Whether these errors are systematic or statistical on our part or on the experimentalists' parts remains to be seen. All of the input parameters are obtained from other kinetic and thermodynamic experiments and are based on the framework of RFOT theory. There are no adjustable parameters in these comparisons.

\subsection{Results: Calorimetry at Different Cooling Rates}

\begin{figure}
\includegraphics[width=19pc,angle=0]{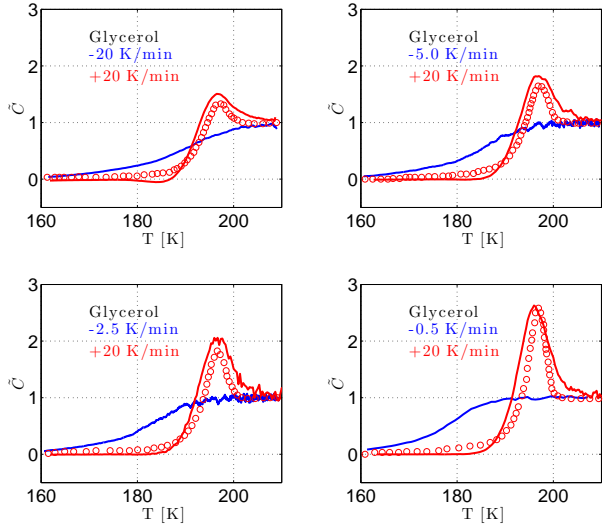}%
\caption{ (Color online) Comparison with heat capacity data obtained through DSC. Experimental data are depicted by red circles for heating scan. Red and blue lines represent heating and cooling curves obtained from our calculation. \label{fig:Gly_ChangeCoolingRate}}
\end{figure}

Figure~\ref{fig:Gly_ChangeCoolingRate} illustrates DSC measurements recorded with various cooling rates, for a single material that is reheated from several lower temperature states and compare them with numerical results simulated using the experimental cooling and heating rates. The experimental results on glycerol were obtained by Wang \textit{et al.}\cite{wang2002ScanningCalorimetry} The four scans differ in the cooling rate (-20, -5, -2.5, -0.5 K/min) and only the up-scans at a common heating rate of +20 K/min are reported. As can be seen, a slower cooling rate yields a more prominent peak in the heat capacity on heating scan and the glass transition temperature $T_g$ shifts toward lower temperature. Clearly our continuum equations also produce correct quantitative behavior for nonstandard cooling and heating protocols. Again no parameters are available for adjustment in the RFOT predictions.

\section{Aging and two equilibration mechanisms in glasses}

 Within the RFOT theory the mosaic structures which correspond to local free energy minima give rise to dynamical heterogeneity both in the supercooled liquid and the glass. In an equilibrium system, the statistics of the energies in the local energylandscape libraries of the sample shows the resulting cutoff Gaussian distribution for activated barrier. However, in an aged sample where the system has already fallen out-of-equilibrium, this statistics have to be determined self-consistently by the dynamics of the system and by its detailed past thermal history.\cite{lubchenko2004theory} If the spatial structure is neglected, Lubchenko and Wolynes suggest this can be approximated by the NMT phenomenology. They also predicted deviations from that phenomenology for certain temperature histories. 
 
As we have seen in the RFOT treatment complex spatiotemporal structures develop during the process of aging glasses and rejuvenating glasses. When the system is cooling, the least stable local regions become replaced by regions equilibrated to the ambient temperature $T$. Thus the mosaic now contain patches of  distinct well equilibrated cells and poorly equilibrated cells in terms of fictive temperature. In this way, the aging glass becomes more stable and is more inhomogeneous in its energy distribution than it was initially when it first fell out of equilibrium.\cite{wolynes2009spatiotemporal} Lubchenko and Wolynes thus argued that RFOT theory implies there would be additional heterogeneity to be found at intermediate times which would lead to additional ultraslow relaxations. Such extra relaxations have sometimes been reported in aging studies. \cite{Miller1997Ultraslow,Boucher2011EnthalpyRecovery,Cangialosi2013TwoEquilibration} The statistics in this case of very deep quenched system are  distinguished by the fast patches and the slow patches that have reconfigured first. Lubchenko and Wolynes pointed out that a clearly two-peaked distribution of local energy will arise in such a situation. 
 
To test these ideas quantitatively we performed numerical simulations of the coupled field equations to investigate the statistics of local energy of a polystyrene glass ($M_n = 85\ \text{kg/mol}$, $T_g=375\ K$) aged at temperatures significantly below the glass transition temperature This system has recently been studied experimentally.\cite{Cangialosi2013TwoEquilibration} The samples were first simulated at high temperature ($T_g+10$) for long time. Then the samples were cooled down at a cooling rate of 20 K/min to reach 350 K, followed by stabilization at the aging temperature $T_a$.	

\begin{figure}
\includegraphics[width=19pc,angle=0]{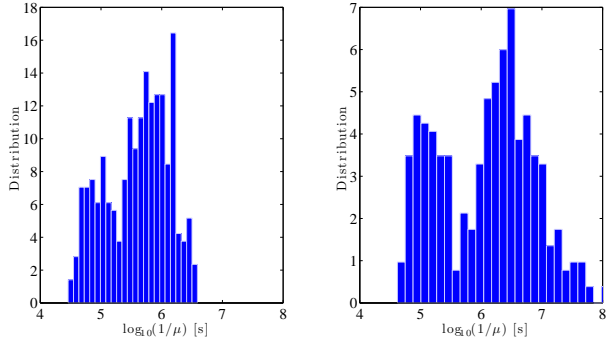}%
\caption{ (Color online) Bimodal distribution of logarithm of relaxation times which represents local energy. The samples were equilibrated at high temperature, then cooled down to a significantly low temperature and aged at the aging temperature $T_a=365\ K$ (left) and $T_a=363\ K$ (right).  \label{fig:bimodal_log_tau}}
\end{figure}

Figure \ref{fig:bimodal_log_tau} shows the distribution of the logarithm of the relaxation times (the inverse of the mobility field roughly equivalent to the local energy) after the samples were stabilized at low temperature $T_a$. The fast peak of the relaxation times corresponds to the regions which have not yet been equilibrated to the new ambient temperature. These fast regions are relatively unstable regions. The slow peak appears due to the regions that originally were less stable and that now have already reconfigured themselves to the new ambient temperature and are now much more stable than those left unequilibrated. The ultraslow peak time is seen to be consistent with the equilibrium relaxation time, i.e. it follows the Vogel-Fulcher-Tammann equation. Obviously the bimodal statistics of the aged glass is different from the mostly single peaked distributions found in equilibrium supercooled liquids. Our simulation results agree quite well with recent experimental and detailed observations by Cangialosi \textit{et. al.} \cite{Cangialosi2013TwoEquilibration} on the polystyrene glass. The separation of the two peaks will become even more pronounced at lower temperatures. Numerical calculations for samples stabilized at still lower temperature would require considerably longer computational times than we had available. While the dichotomic nature of the aged glass structure shows up sharply in relaxation experiments it may also show up as mild density fluctuations that might be detected by scattering or scanning probes.

\begin{figure}
\includegraphics[width=19pc,angle=0]{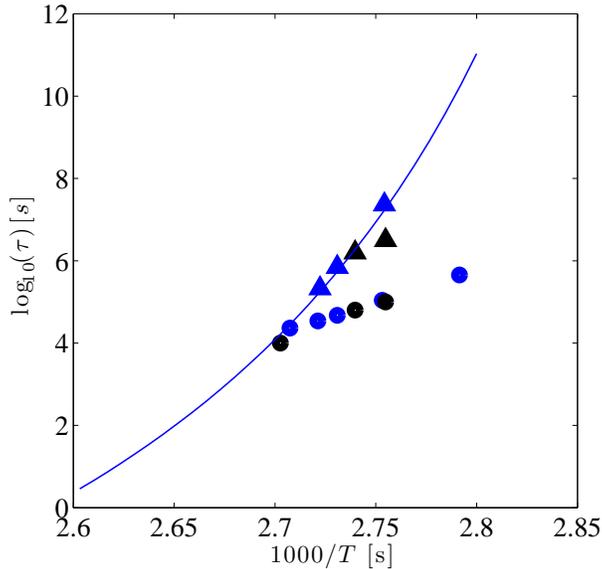}%
\caption{ (Color online) Logarithm of relaxation times corresponding to the first (circle) and the second (triangle) peaks of the distribution of the relaxation times. Blue markers denote experimental data \cite{Cangialosi2013TwoEquilibration} and black markers are simulation results.  \label{fig:compare_equilibration_times}}
\end{figure}

\section{Conclusion and discussion}
In summary we have derived from an extended mode coupling theory the continuum equations for fluctuating mobility and fictive temperature fields in which the activated dynamics from RFOT theory is included. This derivation forms a bridge between mode-coupling theory and the quasistatic aspects of the random first order transition theory of glasses. This formalism allows us to study dynamical heterogeneity and calorimetric measurements in aging glasses in quantitative detail.

It should be possible to apply the same approach to other phenomena where glassy dynamics couples to other fields in a glass. For example the shear banding observed when a system is under the influence of external stress/strain may be studied by coupling the mobility to the deformation of the system using elastic theory and mechanical flow equations.\cite{Hays2000Microstructure} The coupled diffusion may give overshoot phenomena like those seen in some of the experiments perhaps explaining the sensitivity to impurities.\cite{Rossi1995Phenomenological} In polymer glasses, adding a field describing the chemical kinetics of forming bonding constraints to the description of mobility field may elucidate a more complete theory of the chemical aging of those polymeric glasses.\cite{zhao2013using}
\begin{acknowledgments}
Financial support by the D.R. Bullard-Welch Chair at Rice University and a Royal Thai Government Scholarship to A.W. are gratefully acknowledged.
\end{acknowledgments}

\appendix
\section{Derivation of full relaxation formula} \label{app:appendix1}
The free energy profile for conversion from a nonequilibrium initial state to the other usually equilibrium state is\cite{lubchenko2004theory}
\begin{eqnarray}
F(N) = [f_{eq}(T) - \phi_{in}(T)]N + \gamma N^x. \label{eq:A1}
\end{eqnarray}
where $f_{eq}(T)$ is the total bulk free energy per particle of the final state at temperature $T$, $\phi_{in}(T)$ is the internal free energy per particle of the initial state, $\displaystyle \gamma = \frac{3 \sqrt{3 \pi}}{2} k_B T_K \ln \left[ \frac{(a/d_L)^2}{\pi e} \right]$ is the surface tension, $a$ is an interparticle spacing, $d_L$ is the mean square fluctuations of a particle in a given basin.\cite{Stillinger1982HiddenStructure} Accounting for the wetting effect,\cite{villain1985equilibrium} the mismatch exponent is $x = 1/2$. When this exponent is used in Eq.~(\ref{eq:A1}), we obtain the most probable rate:
\begin{eqnarray}
\bar \mu = \mu_0 \exp \left(- \frac{\gamma^2}{4 k_B T [\phi_{in} - f_{eq}]} \right). \label{eq:A2}
\end{eqnarray}

When the initial state is at equilibrium, the free energy driving force $f_{eq} - \phi_{in}$ equals$ - T s_c$. If the liquid is quenched and aging, the driving force $f_{eq} - \phi_{in}$ is negative. Upon cooling initial glass state is trapped at $T_g$, $\phi_{in}(T) = \phi(T_g) \equiv \phi_g$. The free energy of liquid state at an equilibrium temperature $T$ is:
\begin{eqnarray}
 f_{eq}(T) & = & \phi_K - \int_{T_K}^{T} dT' S_c (T') \nonumber \\
	     & = & \phi_K - \Delta c_p(T_g) T_g \left( \frac{T - T_K}{T_K} - \ln \frac{T}{T_K} \right) \nonumber
\end{eqnarray}
where we have used Angell's empirical form for the configurational entropy, $s_c(T) = \Delta c_p(T_g) T_g (1/T_K - 1/T)$ and $\Delta c_p(T) = \Delta c_p(T_g) (T_g/T)$. Noting that, at fictive temperature $T_f$, $f_{eq}(T_f) = \phi_f - T_f s_c(T_f) = \phi_f - \Delta c_p(T_g) T_g \left( \frac{T_f - T_K}{T_K} \right)$ gives the ideal glass state energy $\displaystyle \phi_K  = \phi_f -\Delta c_p(T_g) T_g \ln \frac{T_f}{T_K}$. The free energy driving force is:
\begin{eqnarray}
 f_{eq}(T) - \phi_{in}(T_f)  & = & -\Delta c_p(T_g) T_g \ln \frac{T_f}{T_K}  \nonumber \\
 & \ & - \Delta c_p(T_g) T_g \left( \frac{T - T_K}{T_K} - \ln \frac{T}{T_K} \right) \nonumber \\
 & = & - \Delta c_p(T_g) T_g \left( \frac{T - T_K}{T_K} - \ln \frac{T}{T_f} \right). \label{eq:feqphiin} \nonumber
\end{eqnarray}

Substituting the above equation into Eq.~(\ref{eq:A2}), the full relaxation formula is given in terms of the ambient temperature $T$ and the fictive temperature $T_f$
\begin{eqnarray}
\bar \mu (T,T_f) = \mu_0 \exp \left( - \frac{\gamma^2}{4 k_B T \Delta c_p Tg \left[ \frac{T-T_K}{T_K} - \ln{\frac{T}{T_f}} \right] } \right) \nonumber
\end{eqnarray}
where $\displaystyle \gamma = \frac{3 \sqrt{3 \pi}}{2} k_B T_K \ln \left[ \frac{(a/d_L)^2}{\pi e} \right]$.

\section{Derivation of correlation length in aging regime} \label{app:corrlength}
The activation free energy profile is written as

\begin{equation}
     F(N) = [f_{eq}(T) - \phi_{eq}(T)]N + \gamma N^x \label{eq:B1}
\end{equation}
where the surface energy is equal to $\displaystyle \gamma = \frac{3 \sqrt{3 \pi}}{2} k_B T_K \ln \left[ \frac{(a/d_L)^2}{\pi e} \right]$.

Using the mismatch exponent $x=1/2$ in Eq.~(\ref{eq:B1}) we must solve for the $N^*$ where $F(N^*) = 0$. For $T<T_g$, the number of particles in the  cooperatively rearranging molecular structure is:
\begin{eqnarray}
 N^* & = & \frac{ \gamma_0^{2}  }{ [-(f_{eq}(T) - \phi_{in}(T))]^{2}}. \label{eq:Nstar} 
\end{eqnarray}

Substituting $f_{eq}(T) - \phi_{in}(T)$ from Eq.~(\ref{eq:feqphiin}) into Eq.~(\ref{eq:Nstar}), we obtain $N^*$
\begin{eqnarray}
 N^* & = & \frac{ \left(  \frac{3 \sqrt{3 \pi}}{2} k_B T_K \ln \left[ \frac{(a/d_L)^2}{\pi e} \right]   \right)^{2}  }{ \left[\Delta c_p(T_g) T_g \left( \frac{T - T_K}{T_K} - \ln \frac{T}{T_f} \right) \right]^{2}}. \nonumber
\end{eqnarray}
We can write
\begin{eqnarray}
 N^* = \left(\frac{\xi}{a} \right)^3 = \frac{\left(  \frac{3 \sqrt{3 \pi}}{2} k_B T_K \ln \left[ \frac{(a/d_L)^2}{\pi e} \right]   \right)^{2}  }{ [\Delta c_p(T_g) T_g]^{2}} \cdot \frac{1}{ \left( \frac{T - T_K}{T_K} - \ln \frac{T}{T_f} \right)^{2}}  \nonumber
\end{eqnarray}
In the equilibrium case $T_f \rightarrow T$, so the $\ln(T/T_f)$ term vanishes and we recover the correlation length given by Xia and Wolynes. \cite{xia2000fragilities}

\begin{eqnarray}
 N^* = \left(\frac{\xi}{a} \right)^3 = \frac{\left(  \frac{3 \sqrt{3 \pi}}{2} k_B T_K \ln \left[ \frac{(a/d_L)^2}{\pi e} \right]   \right)^{2}  }{ [\Delta c_p(T_g) T_g]^{2}} \cdot \frac{1}{ \left( \frac{T - T_K}{T_K}  \right)^{2}}.  \nonumber
\end{eqnarray}


\end{document}